\documentclass[prd,reprint, nofootinbib,amsmath,amssymb, aps, floatfix]{revtex4-1}
\usepackage{amssymb,amsmath,amsthm,graphicx,ulem}
\usepackage{hyperref}
\usepackage{graphicx,subfigure}
\usepackage{epsfig}
\usepackage{amsfonts}
\usepackage[usenames]{color}
\usepackage{enumerate}
\usepackage[T1]{fontenc}
\usepackage{float}
\usepackage{xcolor}
\newcommand{\be}{\begin{equation}}
\newcommand{\ee}{\end{equation}}

\begin{document}
\normalem
\title{Joseph Polchinski: A Biographical Memoir}
\author{Raphael Bousso}
 \email{Correspondence: bousso@lbl.gov}
\affiliation{Center for Theoretical Physics and Department of Physics\\
University of California, Berkeley, CA 94720, USA}
\affiliation{Lawrence Berkeley National Laboratory, Berkeley, CA 94720, USA}
\author{Fernando Quevedo}
\affiliation{DAMTP, Centre for Mathematical Sciences\\ Wilberforce Road, Cambridge, CB3 0WA, U.K.}
\author{Steven Weinberg}
\affiliation{Theory Group, Department of Physics, University of Texas, Austin, TX 78712, USA}
\begin{abstract}
  Joseph Polchinski (1954-2018), one of the the leading theoretical physicists of the past 50 years, was an exceptionally broad and deep thinker. He made fundamental contributions to quantum field theory, advancing the role of the renormalization group, and to cosmology, addressing the cosmological constant problem. Polchinski's work on D-branes revolutionized string theory and led to the discovery of a nonperturbative quantum theory of gravity. His recent, incisive reformulation of the black hole information paradox presents us with a profound challenge. Joe was deeply devoted to his family, a beloved colleague and advisor, an excellent writer, and an accomplished athlete.
\end{abstract}

\maketitle

\section{Introduction}

Joseph Polchinski ranks among the greatest theoretical physicists of his generation. His interests were centered on particle physics and quantum gravity, particularly string theory, and he made epochal contributions in these areas. (Unusually for a theorist, his most revolutionary works came between the ages of 40 and 60.) But he was uncommonly broad. His work substantially impacted a range of fields from cosmology to condensed matter physics. 

Polchinski was the very opposite of the caricature of the narrowly focussed theorist, who explores abstruse mathematical structures in an out-of-touch quest for elegance and beauty. He was a full-blooded physicist who cared about understanding Nature, by whatever means he could muster. He was heard saying ``I'm a theoretical physicist first, string theorist second.'' (At the time, shortly after the second superstring revolution that his work had largely triggered, this was not a universal sentiment in the elated community.) As a great pianist is more than a dazzling virtuoso, Polchinski's technical prowess allowed him to perform the most challenging calculations, but always in service of a deeper vision, and in pursuit of a more profound understanding of how the universe works.

When Polchinski came of age as a physicist, the Standard Model of particle physics had recently been worked out. It explained all observed forces but gravity, within the framework of quantum mechanics. The great remaining task since then has been to unify gravity and quantum mechanics. It takes some courage to decide to work on quantum gravity: one expects that the final theory will be simple only at energies and distances that are far out of reach technologically. Many reasonable hypotheses can be probed at best indirectly, through subtle effects. 

Against these odds, Joe made profound and lasting contributions to science. From our early (and personal) vantage point,\footnote{We urge the reader to study Polchinski's own recollections~\cite{Polchinski:2017vik}, which are truly fascinating. They are refreshingly and poignantly honest about the difficulties he faced in his life and career. As such they can be a valuable resource for all of us, and especially to young people entering the field. Other excellent obituaries and tributes written by physicists include Refs.~\cite{obituary, obituary2, obituary3, matt}.} Joe's most significant works include the ``string theory landscape'' as a solution of the cosmological constant problem, and his contributions to the renormalization group and to the black hole information paradox. And through the breadth and depth of its impact, Polchinski's 1995 discovery of D-branes truly stands out.

D-branes mark a watershed in theoretical physics.
Before D-branes, string theory was formulated and understood largely at a perturbative level, as a sum over diagrams. The theory offered a promising approach to quantum gravity---one could compute how gravitons scatter---and to unification, with the hope of deriving the observed symmetries, forces, and particles, from the rigid structure provided by the theory.

But string theory could not be applied to strongly gravitating systems such as black holes, or to cosmology. Thus, many pressing questions could not be addressed: what is the origin of black hole entropy? What happened at the big bang? Polchinski was the first to recognize that the symmetries of string theory require the existence of D-branes, extended objects that are intrinsically non-perturbative.

This insight led to dramatic progress on several deep and long-standing problems. Polchinski took the lead in many of these developments. His D-brane paper \cite{Polchinski:1995mt} and his nearly simultaneous work with Witten \cite{Polchinski:1995df}\, form cornerstones of the ``second superstring revolution,'' which unified what were then five distinct versions of string theory.  

We see the impact of Polchinski's discovery in the quality of the work it elicited in other leading physicists. Strominger and Vafa~\cite{Strominger:1996sh} succeeded in describing certain black holes microscopically as stacks of D-branes. They were able, for the first time, to compute the black hole entropy statistically, matching the thermodynamic formula of Bekenstein and Hawking which is considered one of the greatest successes of string theory. Within a year, Banks, Fischler, Shenker and Susskind~\cite{Banks:1996vh} formulated Matrix theory, using a large number of D0-branes to provide the first concrete proposal for a nonperturbative description of string theory (often referred to as M-theory). In 1997, Maldacena formulated his celebrated Anti-de Sitter/Conformal Field Theory conjecture~\cite{Maldacena:1997re}, providing a complete, nonperturbative definition of a quantum theory of gravity in a large class of spacetimes. The insight that we might live on a D-brane~\cite{ArkaniHamed:1998rs, Randall:1999ee, Ibanez:2012zz} opened vast new directions in particle phenomenology. And finally, D-branes and their associated fluxes underlie the string landscape~\cite{Bousso:2000xa,Giddings:2001yu,Kachru:2003aw}, as we discuss below in the context of the cosmological constant problem.

D-branes were not mere catalysts of these developments. Rather, they play a central role in the new fields that formed in their wake. They have become a ubiquitous tool, nearly as familiar to today's graduate students as quantum fields. Their applications range far beyond particle physics and quantum gravity, to problems such as superconductivity and the quark-gluon plasma.

\section{Early life}

Joseph Gerard Polchinski Jr.\ was born in White Plains, N.Y., on May 16, 1954, to Joseph and Joan Polchinski (n\'ee Thornton). He grew up there and in Tucson, Arizona, developing an early fascination with math and physics. He went to Caltech, where his interests included devising pranks with his friend, the physicist Bill Zajc, as well as taking classes with Richard Feynman and Kip Thorne~\cite{Polchinski:2017vik}.

For graduate school, Joe picked Berkeley, partly out of his love of California, partly because of a Hertz fellowship. He worked with Stanley Mandelstam on quark confinement and---unsurprisingly, given the difficulty of the problem---published nothing at all in his entire graduate career. In his ``Memories,'' Joe blames his own ``lack of common sense and of any collaborative instinct.''

We deliberately keep this section short, in the hope that it will provoke readers to read Joe's ``Memories'' \cite{Polchinski:2017vik}. They are a fascinating, concise, and beautifully written account of his life and work. Below we offer our perspective on Polchinski the physicist.

\section{Early career and first important results}

Having published zero papers as a graduate student was not the academic death sentence it might be today. Joe went on to SLAC for his first postdoc. There he worked with Leonard Susskind, who would become an important lifelong influence.

Joe and Lenny studied several aspects of supersymmetry. During the early 1980's several groups had realized that the special quantum behaviour of supersymmetric field theories could play an important role in physics beyond the Standard Model. The Lagrangian of the simplest supersymmetric gauge theories $\cal{N}=1$ consists of two terms, denoted F and D terms. F-terms were already known not to be affected by quantum corrections. In their first article, together with Willy Fischler, Peter Nilles and Stuart Raby \cite{Fischler:1981zk}, Joe and Lenny proved that the D-terms were also not renormalized by quantum corrections as long as the sum of all the charges of the matter fields is zero (a condition that was later identified as  the absence of gravitational anomalies).

Second, Joe and Susskind~\cite{Polchinski:1982an}  were able to prove that the energy scale of supersymmetry breaking could be much higher than the scale that determines the difference in mass between fermions and bosons (which to address the hierarchy problem was expected to be close to the TeV scale). They presented arguments and detailed calculations to show the stability of scales and the natural  presence of a ``hidden sector'' in supersymmetric extensions of the Standard Model. In the Standard Model, the symmetry breaking sector (the Higgs) couples directly to the matter sector. But in supersymmetric theories, the sector that breaks supersymmetry does not have to couple directly to the observable sector, and the source of supersymmetry breaking can be at much higher energies. These ideas have been behind the subsequent efforts towards gravity mediation in supergravity and string theory constructions.

For his second postdoc at Harvard, Joe started to collaborate with Mark Wise and Luis Alvarez-Gaum\'e \cite{AlvarezGaume:1983gj}. They managed to implement electroweak symmetry breaking in a supersymmetric version of the Standard Model by using the renormalization group running, in the sense that a negative quadratic term for the Higgs field needed for electroweak symmetry breaking is naturally induced by having a very heavy top mass running in the loops. This result (obtained independently by Luis Ib\'a\~nez and Graham Ross \cite{Ibanez:1982fr} ) gave a strong argument in favour of supersymmetric extensions of the Standard Model. This was reinforced by the discovery in the 1990s of the top quark with a very heavy mass, precisely as needed for this radiative electroweak symmetry breaking.

In an important single-author paper while at Harvard \cite{Polchinski:1983gv}\, Joe was able to extend the Wilsonian approach to renormalization to arbitrary cut-offs and to derive a functional equation for the renormalized action. In this regard, he pioneered the development of the ``exact renormalization group''  formalism, in which a single complicated equation embodies all of the usual renormalization group equations to all orders in perturbation theory. He proved that under broad conditions, any theory formulated with an ultraviolet cut-off will flow to a renormalizable theory as the cut-off is decreased.

Joe's arrival at the University of Texas, Austin, essentially coincided with the start of the first string theory revolution.  Joe immediately started playing a leading role in the Weinberg theory group. He wrote a beautiful paper on the one-loop amplitude for the closed bosonic string in the path integral Polyakov formalism \cite{Polchinski:1985zf}. This explicit calculation of the vacuum amplitude emphasized the crucial role of modular invariance, the difference with the field theory spectrum, and the relation to  the finite temperature partition function. It had an enormous impact in the string community, and it led to many further developments both in bosonic and supersymmetric string theory.

\section{D-branes}

In this section, we will focus on the lead-up to Polchinski's discovery of D-branes, as we have already discussed its impact and significance in the introduction.

In 1986, with his students Jim Hughes and Jun Liu~\cite{Hughes:1986fa}, he found the effective action of supersymmetric branes, in an attempt to understand partial supersymmetry breaking ($\mathcal{N}=2\to \mathcal{N}=1$) in supersymmetric field theories. This article directly led to the work of Eric Bergshoeff, Ergin Sezgin and Paul Townsend~\cite{{Bergshoeff:1987cm}} on the maximal 11-dimensional supermembrane, which in turn triggered the early work on general p-branes~\cite{Duff:2018goj}.

In 1989, in a completely different project, Joe sought an improved understanding of T-duality, the string symmetry relating large to small distances. With his students Jin Dai and Robert Leigh, he discovered that T-duality for open strings exchanges Neumann and Dirichlet boundary conditions~\cite{Dai:1989ua}. In the latter case the endpoints of the string had to be fixed at a lower-dimensional surface, which they denoted Dirichlet-brane or {\it D-brane}.\footnote{In this same article the concept and name of {\it orientifold} was introduced. It corresponds to simultaneous twists in target space and the string worldsheet, generalizing the standard concept of orbifold. Related work was done by Augusto Sagnotti \cite{Sagnotti:1987tw} and Petr Ho\v{r}ava  \cite{Horava:1989vt} independently.} Due to the presence of gravity in string theory, these objects are not rigid surfaces but must be dynamical.

In 1992 Joe moved to Santa Barbara as one of the few faculty members of the Institute for Theoretical Physics (now KITP). Three years later he realised the crucial role that D-branes can play in the equivalence among all different string theories, hints of which had begun to emerge by then. In a short paper~\cite{Polchinski:1995mt} he computed the corresponding string amplitude that helps to identify D-branes as solitonic non-perturbative objects of different dimensionalities carrying both tension and charge. He established the generalized Dirac quantization condition, similar to monopoles. D-branes are charged not under electromagnetic field but under more general antisymmetric tensor fields appearing in the spectrum of string theories.

\section{Explaining the Smallness of the Cosmological Constant}

In 1999, Polchinski turned his attention to the problem of the cosmological constant (CC). A nonzero CC was first introduced by Einstein in a futile and misguided attempt to make his general theory of relativity compatible with a presumed static universe. In Einstein's equation, the CC is mathematically equivalent to the energy density of the vacuum, which was later recognized to be an inevitable feature of quantum field theory.\footnote{A more recent name for the CC (aside from ``vacuum energy'') is ``dark energy.'' This terminology encourages us to keep an open mind: perhaps the acceleration of the expansion of the universe is driven by a time-dependent effect. However, such models are strictly more complicated than a pure CC. They fine-tune the CC but contain additional parameters that need to be tuned. In more than two decades since the first measurement of the CC, no observational evidence has emerged that would compel us to consider such complications.} Contributions to the vacuum energy density from Standard Model fields are at least 60 orders of magnitude larger than the value first observed in 1998~\cite{Riess:1998cb, Perlmutter:1998np} (and long-known upper bounds). How come they cancel to such exquisite precision? This is the cosmological constant problem: why is the vacuum energy so tiny?\footnote{For reviews of the CC problem, see Refs.~\cite{Weinberg:1988cp,Polchinski:2006gy,Bousso:2007gp}; the latter two references also summarize the Bousso-Polchinski proposal. For a nontechnical summary, see Ref.~\cite{Bousso:2004fc}. Joe offers his own account of the gestation and impact of this work in Sec.~9.6 of Ref.~\cite{Polchinski:2017vik}: its implications troubled him greatly, but ``one must follow science where it leads'' (Sec.~10.7).}

One of the present authors (SW) had recognized in the 1980s that the unnatural smallness of the CC may be a selection effect~\cite{Weinberg:1987dv}. A very small CC is needed for large structures to form in the universe. Perhaps there are many regions with different values of the CC, and observers find themselves where structure is possible? If so, the CC should not be much smaller than the present matter density. This proposal would have been ruled out by significantly tightened upper bounds on the CC. But instead a positive value was observed in 1998---just what would be expected if the CC was biased by observer selection, among many possible locations in the universe.

Yet, it proved difficult to implement the idea of the CC as a selection effect, at the concrete level of obtaining a theory with suitable vacuum structure and yet with dynamics consistent with the observed big bang cosmology. First, the required number of vacua (long-lived field configurations, each with different CC) would have to exceed $10^{123}$. How would a simple theory give rise to so many different versions of empty space? Secondly, if one simply posits an effective potential with a huge number of minima, the theory runs into sharp conflict with observation: it predicts a universe with far too little matter and radiation. These problems, first articulated by Abbott and others in the 1980s, appeared severe and stood unsolved~\cite{Abbott:1984qf,Brown:1988kg}.

In Polchinski's work with one of the present authors (RB)~\cite{Bousso:2000xa}, a key insight was to think of D-branes as building blocks of empty space. String theory has 9 spatial dimensions, so our observation of only 3 requires that 6 are ``compactified'' and too small to observe. There were many ways one could envisage doing this, but all known approaches would lead to new long-range fields or to a negative value of the CC, neither of which are observed. Polchinski noticed that the presence of D-branes and their associated fluxes in the extra dimensions would have profound implications. They would make it possible to avoid unwanted light fields (this was corroborated by later work of Kachru {\em et al.} as we will see next), and to obtain a vacuum such as ours in string theory. 

Moreover, D-branes and fluxes greatly increase the number of possible ways to make what a low-energy observer would perceive as empty space. Simple combinatorics yields an enormous number of vacua and hence, a dense spectrum of the CC. This solved the first challenge in implementing Weinberg's approach. 

Because fluxes can wrap different topological cycles, one obtains a high-dimensional effective potential (a ``landscape''). Neighboring vacua can have enormously different vacuum energy, yet all the vacua can be populated by the cosmological dynamics. When one vacuum decays, matter and radiation can be produced in the region containing the new vacuum. This addressed the second challenge, allowing for a hot big bang.

Whether a large landscape of vacua is the correct solution to the CC problem is not known. The vacuum structure of Nature, like any other scientific question, must be settled by observation. We lack the technology to probe our vacuum at the required energies, and our theoretical control is not yet strong enough to devise decisive indirect tests. 

But it is significant that after more than 40 years of serious attempts at the CC problem, Polchinski's insights underly the only known viable approach.  If the landscape proves correct, it is a revolution. The particles and forces we observe would not be unique. This challenges naturalness as the dominant paradigm of particle physics: selection effects, rather than symmetries, may explain not only the CC but also other parameters such as masses and couplings. (Indeed, a natural origin of the scale of the weak interaction is already in tension with and may eventually be ruled out by present and future accelerators.) In cosmology, the big bang would correspond to a phase transition, and the visible universe to one possible outcome of a chain of such transitions.

\section{Moduli Stabilization}

Stabilizing the size and shape of the extra dimensions has been a major challenge for higher dimensional theories since the original work of Kaluza and Klein in the early 20th century. Among the most influential theoretical articles of the first decade of this century is ``Hierarchies from fluxes in string compatifications''  by Giddings, Kachru, and Polchinski (GKP)~\cite{Giddings:2001yu}. This article established how fluxes of antisymmetric tensor fields can stabilize many of the  different moduli (shape and size of the extra dimensions), as anticipated in Ref.~\cite{Bousso:2000xa}. GKP left only a class of moduli known as K\"ahler moduli unfixed, but with the interesting outcome that they break supersymmetry while still keeping vanishing cosmological constant at tree-level, a property known as no-scale supergravity.

This work triggered the subsequent work of Kachru, Kallosh, Linde and Trivedi~\cite{Kachru:2003aw} (KKLT) that added non-perturbative effects and antibranes to also fix the K\"ahler moduli and uplift the minimum to de Sitter space. 

The GKP construction is very explicit, illustrating how three-form fluxes fix the dilaton and complex structure moduli (sizes of three-cycles) with a concrete description both from 10 dimensions and from the four dimensional effective actions. This construction underlies the string theory landscape~\cite{Bousso:2000xa}, in the sense that the source of the large number of metastable solutions are precisely the fluxes in the GKP part of  the full GKP and KKLT constructions. The fluxes are integers due to a Dirac quantization condition but several integers can take a range of values, leading to an enormous number of possible combinations. These properties were anticipated by Bousso and Polchinski in their approach to the cosmological constant problem. The GKP and KKLT constructions lend substantial support to the BP proposal, by providing an explicit string theory realization of the scenario. 

Independently of the cosmological constant problem, moduli stabilization removed a major stumbling block for string theory to make contact with low-energy physics. (Massless moduli fields would mediate long range interactions, contrary to observations.) Moreover, the GKP flux compactifications naturally induce warping in the extra dimensions which explicitly realize the popular Randall-Sundrum scenario in string theory. In this sense the warping introduces a natural way to create hierarchies that may connect the Planck scale to smaller scales such as the TeV scale governing the Standard Model.

The GKP and KKLT constructions influenced thousands of articles addressing concrete phenomenological and cosmological questions  of string compactifications.  In one proposal, collisions of Polchinski's D-branes and anti-branes may give rise to cosmological inflation~\cite{Dvali:1998pa, Burgess:2001fx}. The annihilation process can also produce lower dimensional branes such as relic cosmic strings that could give rise to observational signatures~\cite{Burgess:2001fx,Sarangi:2002yt}. Joe's deep physical intuition immediately entered in action and explored its consequences together with Edmund Copeland and Rob Myers~\cite{Copeland:2003bj}. Potential observation of cosmic strings is one of the very few opportunities to test string theoretical ideas at very high energies. 

Moduli stabilization and its potential physical implications is still a very active research area. In one of his very last projects Joe managed to solve a challenge regarding an obstruction to  consistently have anti-branes in the presence of fluxes that give rise to de Sitter space in KKLT~\cite{Polchinski:2015bea}. This settled a long debate in the subject.

\section{Black Holes, Information, and Firewalls}

The history of the black hole information paradox is full of unexpected turns and surprises. Jacob Bekenstein's 1972 conjecture~\cite{Bekenstein:1972tm} that black holes carry entropy was at first rejected by Stephen Hawking~\cite{Bardeen:1973gs}. Ironically, Hawking's objection was that Bekenstein's entropy would imply a nonzero temperature (by the First Law), and hence would require black holes to radiate thermally. This seemed absurd, because classical black holes cannot emit anything. But in an unrelated calculation, Hawking soon found that black holes do radiate, by a quantum effect~\cite{Hawking:1974sw}. When no infalling matter compensates, this implies that a black hole eventually loses all of its mass and disappears. It will then have returned the corresponding amount of energy to the exterior in the form of a Hawking radiation cloud. 

Hawking was able to show, moreover, that the quantum state of the radiation would be {\em mixed} (roughly, thermal)~\cite{Hawking:1976ra}. Thus the final state would be essentially unique, in that it would depend only on the mass and angular momentum of the matter that made the black hole. It would not otherwise depend on its initial quantum state.

This was a shocking claim. It meant that the unitarity of quantum mechanics would break down in the presence of a black hole. Normally, one can use the Schr\"odinger equation to evolve a quantum state forward or backward in time. The state at a different time is computed by the action of a unitary operator that depends on the dynamics. This map is one-to-one, allowing for perfect predictability. For large systems, this remains true at least in principle, if not in practice due to limited control. But for black holes, Hawking found that unitarity would be false even as a matter of principle. Any pure state resulting in a black hole of a given mass and angular momentum would be mapped to the same mixed state, that is, to a classical probability distribution over pure states.

Hawking's claim disturbed many physicists, as it challenged a core principle of quantum mechanics. But no error was found in his calculation, and his assumptions seemed innocuous: mainly, that the horizon of a large black hole, when no matter is falling in, is just like empty space everywhere else. This is a straightforward consequence of Einstein's general theory of relativity. If this assumption were wrong, Einstein's theory would seem to break down completely. It would fail in a regime where it was expected to be arbitrarily accurate, namely far from the quantum domain, when the curvatures of space and time are small.

In the early 1990s, Gerard 't Hooft, Leonard Susskind, John Preskill and others came up with an unorthodox attempt to thread the needle, hoping to save quantum mechanics without breaking general relativity. This idea became known as black hole complementarity~\cite{Susskind:1993if}, as it assigned two ``complementary'' descriptions to the black hole. Viewed from the outside, it would behave like any other object, with some actual structure at the horizon that would preserve information about the quantum state, in harmony with the quantum mechanical evolution tested by this observer. Yet, an observer freely falling across the horizon would experience nothing special, just empty space, as dictated for that observer by general relativity.  

Naively these two descriptions conflict. But complementarity cleverly exploited some undeniable facts about black holes. By definition, an observer inside the black hole cannot send a signal to the exterior, nor could she return to the outside to check whether or not information was lost. Perhaps more surprisingly, one can also check that an observer on the outside cannot send a signal to a friend (say, that the information came out), if the latter entered a black hole too long ago. These obstructions seemed to conspire against any possibility of verifying the apparent differences in their description of the black hole. It thus appeared that no contradictions between general relativity and quantum mechanics would be encountered in any conceivable experiment involving large black holes, so long as the experiment itself did not violate known laws of physics. 

Though some physicists (notably, Samir Mathur) were vocally skeptical about its viability, complementarity is where things stood for nearly 20 years. But in 2012, Polchinski (working with A.~Almheiri, D.~Marolf, and J.~Sully) found a sharp contradiction in the framework~\cite{Almheiri:2012rt}. Complementarity was not enough to resolve the black hole information paradox. The AMPS collaboration accomplished this by taking advantage of some of the counterintuitive properties of quantum information. They showed that the joint assumptions of unitarity and empty space at the horizon would allow for the construction of an impossible quantum state in the laboratory of an observer entering the black hole. One of these assumptions (or else the validity of effective field theory outside the black hole) would have to be abandoned.

If the AMPS paper had appeared 30 years earlier, it might have convinced many physicists of Hawking's claim that information is lost. Since Hawking's original work, however, the theoretical prior had shifted significantly towards unitarity. The AdS/CFT correspondence defined a compelling quantum gravity theory in a setting where the formation and evaporation of black holes can occur, and it simply did not allow for information to be lost.

With no wiggle room for a compromise like complementarity, AMPS argued that the most conservative conclusion was a complete breakdown of general relativity at the horizon. But what replaces empty space? With minimal assumptions, AMPS showed that all quantum fields would be excited at arbitrarily small distance scales near the horizon, and hence at very high energies. Instead of empty space, the boundary of a black hole would be a ``firewall.''

Firewalls came as an absolute shock to many in the quantum gravity community. It took a mind like Joe's to cut through the preconceptions and return us to the stark choice between information loss and a total breakdown of Einstein's theory. Something big has to give, so the firewall crisis is likely to contribute to a major leap in our understanding of Nature. An army of physicists is sharpening its tools, hoping to determine how to get rid of firewalls, or else to understand how they would form and persist. We are developing powerful new methods and applying ever more sophisticated techniques from quantum information theory to the study of spacetime and gravity. 

\section{The Broad Theorist}

Even though most of Joe's efforts concentrated in the development of string theory, he was a theoretical physicist in the broadest sense. Many theorists are at home only in a very narrow range of highly mathematical topics. Polchinksi was different. He did work of real importance in a remarkable variety of fields. 

With Fischler, Polchinski challenged what seemed to be a breakthrough by Sidney Coleman and Stephen Hawking in understanding the cosmological constant. They had claimed that quantum fluctuations produce a probability distribution for the vacuum energy that is sharply peaked at value zero, but Fischler and Polchinski showed that the approximations used by Coleman and Hawking cannot be trusted. Furthermore, together with Fischler and Morgan \cite{Fischler:1990pk}, he developed a general formalism to study vacuum transitions through bubble nucleation in semiclassical gravity, generalising and putting on firmer grounds previous work of Coleman and De Luccia as well as of the general ideas of Farhi, Guth and Guven on the possibility of creating a universe in the laboratory. 

Polchinski pioneered several applications of the renormalization group to quantum field theory and, as mentioned above, made fundamental contributions to the formulation of the renormalization group. He also wrote a beautiful article on the effective field theory approach to describe Fermi liquids, an elegant contribution to condensed matter physics inspired by general ideas of effective quantum field theories. Using his own words, Joe proved that BCS superconductivity was due to asymptotic freedom just as the confining of quarks. This article is becoming more relevant today, as the condensed matter and high energy communities are finding closer common research interests. He has also several articles on the applications to condensed matter of the AdS/CFT correspondence, and much earlier a pure condensed matter article with Fisher and Kane \cite{Kane:1994vb}\, on the quantum Hall effect.

``Polchinski's paradox'' refers to an obstacle he formulated to a proposal by Kip Thorne and collaborators: the possibility of  using wormholes to travel in time. Joe's idea was to throw a billiard ball through a wormhole in such a way that after coming out in the past it hits itself, preventing itself from entering the wormhole. This is better known in the popular science literature \cite{Thorne:1994xa}. 

Polchinski made an important contribution to the foundations of quantum mechanics. He successfully torpedoed a possible generalization of quantum mechanics invented by one of us (SW), by showing that it would allow instantaneous communication at a distance, and even communication between different histories in a many-histories interpretation of quantum mechanics. This work has set an important constraint on any future attempt to go beyond the usual version of quantum mechanics. 

\section{Joe's Legacy}

Joe has left a vast legacy in many respects. As a family man, he is dearly missed by his wife Dorothy and his sons Steve and Daniel. As a great scientist with the deepest mind, tremendous physical intuition, impressive calculational tools, a wide knowledge and a deep understanding of theoretical physics, he was admired and respected by his colleagues. His students enjoyed his unusual gift for combining friendliness and brilliance. 

He dedicated much of his research efforts to string theory. Indeed, he did not shy from publicly defending and promoting work in this field, so in a way he represented ``the voice of string theory.''

But as we mentioned at the outset, Joe always thought of himself as a ``theoretical physicist first, string theorist second.'' (This is a literal quote, to one of us, RB.) This attitude is reflected in the breadth and richness of his knowledge, and in his research achievements spanning cosmology, particle physics, black holes, and condensed matter physics.  

His lectures were masterpieces of clarity, elegance and logical presentation, never exceeding the appropriate number of words. His writing skills were unparalleled. Besides his two major string theory textbooks \cite{books}, he wrote beautiful reviews on many aspects of theoretical physics (see for example\cite{Polchinski:1992ed, Polchinski:1996nb,  Polchinski:2010hw, Polchinski:2014mva, Polchinski:2015pzt, Polchinski:2016xto}). They remain the standard ways to introduce the corresponding field. His  famous ``star'' plot of all string theories being related to each other has been reproduced in many presentations as the standard way to explain the unity of string theories. His introduction of terms such as D-branes, orientifolds, discretuum, etc.\ are part of the daily vocabulary among thousands of theoretical physics worldwide.

For all of us who had the privilege to know him, he has left an indelible mark. He was open enough to listen to everybody independently of their status. He was patient enough to explain physics concepts to less expert colleagues or students. He was kind enough to give respect and dedicate time to everybody.

Joe was a human being living life to its fullest. He loved and obsessed over physical exercise (including estimating in real time the number of calories he was consuming in every meal and decide how much extra exercise he would need to do to compensate). 

Regarding his many qualities as a researcher, one  of his main collaborators, Matt Strassler recalls witnessing Joe's calculational skills and physic intuition~\cite{matt}:
\\

{\it Each calculation was unique, a little gem, involving a distinctive investigation of exotically-shaped D-branes sitting in a curved space. It was breathtaking to witness the speed with which Joe worked, the breadth and depth of his mathematical talent, and his unmatched understanding of these branes.... Somehow his mind was working in places that language does not go, in ways that none of us outside his brain will ever understand. In him, there was something of an oracle.}
\\

And to remember typical expressions that all of us who met Joe will never forget:
\\

{\it Among my favourite memories as a scientist are moments when I taught Joe something he did not know; he would be silent for a few seconds, nodding rapidly, with an intent look -his eyes narrow and his mouth slightly open-  as he absorbed the point.}
\\

We may add to this his tendency to close his eyes while explaining some deep concept. One of his famous anecdotes was that he fell asleep during one of his own lectures in ICTP in 1995 while doing this.

Looking to the future, we can be confident that Joe's imprint will be present in any new discovery about the fundamental understanding of Nature. D-branes could be the basic building blocks of matter; they appear to be the underlying degrees of freedom inside a black hole. They can also be whole universes, and they are the building blocks of the multiverse that we may inhabit.

As Joe once quipped, he discovered both the smallest and the largest things. What better legacy could one hope for?


\begin{thebibliography}{}


  


\bibitem{Polchinski:2017vik} 
  J.~Polchinski,
  ``Memories of a Theoretical Physicist,''
  arXiv:1708.09093 [physics.hist-ph].

\bibitem{obituary}
D.~Marolf and E.~Silverstein,\hfill\\ https://physicstoday.scitation.org/doi/full /10.1063/PT.3.3928

\bibitem{obituary2}
E.~Silverstein,\hfill\\ https://www.quantamagazine.org/joe-polchinski-physicist-who-developed-deep-ideas-and-paradoxes-dies-at-63-20180220/

\bibitem{obituary3}
S.~Carroll,\hfill\\ https://blogs.scientificamerican.com/observations/in-memoriam-joe-polchinski-1954-2018/

\bibitem{matt}
M.~Strassler,\hfill\\ https://profmattstrassler.com/2018/02/05/a-brilliant-light-disappears-over-the-horizon-in-memory-of-joe-polchinski/

\bibitem{Polchinski:1995mt} 
  J.~Polchinski,
  ``Dirichlet Branes and Ramond-Ramond charges,''
  Phys.\ Rev.\ Lett.\  {\bf 75}, 4724 (1995)
  [hep-th/9510017].


\bibitem{Polchinski:1995df} 
  J.~Polchinski and E.~Witten,
  ``Evidence for heterotic - type I string duality,''
  Nucl.\ Phys.\ B {\bf 460}, 525 (1996)
  [hep-th/9510169].


\bibitem{Strominger:1996sh} 
  A.~Strominger and C.~Vafa,
  ``Microscopic origin of the Bekenstein-Hawking entropy,''
  Phys.\ Lett.\ B {\bf 379}, 99 (1996)
  [hep-th/9601029].


\bibitem{Banks:1996vh} 
  T.~Banks, W.~Fischler, S.~H.~Shenker and L.~Susskind,
  ``M theory as a matrix model: A Conjecture,''
  Phys.\ Rev.\ D {\bf 55}, 5112 (1997)
  [hep-th/9610043].


\bibitem{Maldacena:1997re} 
  J.~M.~Maldacena,
  ``The Large N limit of superconformal field theories and supergravity,''
  Int.\ J.\ Theor.\ Phys.\  {\bf 38}, 1113 (1999)
  [Adv.\ Theor.\ Math.\ Phys.\  {\bf 2}, 231 (1998)]
  [hep-th/9711200].


\bibitem{ArkaniHamed:1998rs} 
  N.~Arkani-Hamed, S.~Dimopoulos and G.~R.~Dvali,
  ``The Hierarchy problem and new dimensions at a millimeter,''
  Phys.\ Lett.\ B {\bf 429}, 263 (1998)
  [hep-ph/9803315].


\bibitem{Randall:1999ee} 
  L.~Randall and R.~Sundrum,
  ``A Large mass hierarchy from a small extra dimension,''
  Phys.\ Rev.\ Lett.\  {\bf 83}, 3370 (1999)
  [hep-ph/9905221].


\bibitem{Ibanez:2012zz} 
  L.~E.~Ibanez and A.~M.~Uranga,
  "String Theory and Particle Physics". Cambridge University Press (2012).
  ``String theory and particle physics: An introduction to string phenomenology,''


\bibitem{Bousso:2000xa} 
  R.~Bousso and J.~Polchinski,
  ``Quantization of four form fluxes and dynamical neutralization of the cosmological constant,''
  JHEP {\bf 0006}, 006 (2000)
  [hep-th/0004134].


\bibitem{Giddings:2001yu} 
  S.~B.~Giddings, S.~Kachru and J.~Polchinski,
  ``Hierarchies from fluxes in string compactifications,''
  Phys.\ Rev.\ D {\bf 66}, 106006 (2002)
  [hep-th/0105097].


\bibitem{Kachru:2003aw} 
  S.~Kachru, R.~Kallosh, A.~D.~Linde and S.~P.~Trivedi,
  ``De Sitter vacua in string theory,''
  Phys.\ Rev.\ D {\bf 68}, 046005 (2003)
  [hep-th/0301240].

\bibitem{Fischler:1981zk} 
  W.~Fischler, H.~P.~Nilles, J.~Polchinski, S.~Raby and L.~Susskind,
  ``Vanishing Renormalization of the D Term in Supersymmetric U(1) Theories,''
  Phys.\ Rev.\ Lett.\  {\bf 47}, 757 (1981).


\bibitem{Polchinski:1982an} 
  J.~Polchinski and L.~Susskind,
  ``Breaking of Supersymmetry at Intermediate-Energy,''
  Phys.\ Rev.\ D {\bf 26}, 3661 (1982).


\bibitem{AlvarezGaume:1983gj} 
  L.~Alvarez-Gaume, J.~Polchinski and M.~B.~Wise,
  ``Minimal Low-Energy Supergravity,''
  Nucl.\ Phys.\ B {\bf 221}, 495 (1983).

\bibitem{Ibanez:1982fr}
  L.~E.~Ibanez and G.~G.~Ross,
  ``SU(2)-L x U(1) Symmetry Breaking as a Radiative Effect of Supersymmetry Breaking in Guts,''
  Phys.\ Lett.\  {\bf 110B} (1982) 215.

\bibitem{Polchinski:1983gv} 
  J.~Polchinski,
  ``Renormalization and Effective Lagrangians,''
  Nucl.\ Phys.\ B {\bf 231}, 269 (1984).


\bibitem{Polchinski:1985zf} 
  J.~Polchinski,
  ``Evaluation of the One Loop String Path Integral,''
  Commun.\ Math.\ Phys.\  {\bf 104}, 37 (1986).


\bibitem{Hughes:1986fa} 
  J.~Hughes, J.~Liu and J.~Polchinski,
  ``Supermembranes,''
  Phys.\ Lett.\ B {\bf 180}, 370 (1986).

  
\bibitem{Bergshoeff:1987cm}
  E.~Bergshoeff, E.~Sezgin and P.~K.~Townsend,
  ``Supermembranes and Eleven-Dimensional Supergravity,''
  Phys.\ Lett.\ B {\bf 189} (1987) 75.

\bibitem{Duff:2018goj} 
  M.~J.~Duff,
  arXiv:1812.11658 [hep-th].
  
\bibitem{Dai:1989ua} 
  J.~Dai, R.~G.~Leigh and J.~Polchinski,
  ``New Connections Between String Theories,''
  Mod.\ Phys.\ Lett.\ A {\bf 4}, 2073 (1989).
  
  
  \bibitem{Sagnotti:1987tw}
  A.~Sagnotti,
  ``Open Strings and their Symmetry Groups,''
  in {\em Cargese 1987, Proceedings, Nonperturbative Quantum Field Theory}, 521-528; and Rome II Univ., ROM2F-87-025; [hep-th/0208020].
  
  \bibitem{Horava:1989vt}
  P.~Horava,
  ``Strings on World Sheet Orbifolds,''
  Nucl.\ Phys.\ B {\bf 327} (1989) 461.

\bibitem{Riess:1998cb} 
  A.~G.~Riess {\it et al.} [Supernova Search Team],
  ``Observational evidence from supernovae for an accelerating universe and a cosmological constant,''
  Astron.\ J.\  {\bf 116}, 1009 (1998)
  [astro-ph/9805201].

\bibitem{Perlmutter:1998np} 
  S.~Perlmutter {\it et al.} [Supernova Cosmology Project Collaboration],
  ``Measurements of $\Omega$ and $\Lambda$ from 42 high redshift supernovae,''
  Astrophys.\ J.\  {\bf 517}, 565 (1999)
  [astro-ph/9812133].
  
\bibitem{Weinberg:1988cp} 
  S.~Weinberg,
  ``The Cosmological Constant Problem,''
  Rev.\ Mod.\ Phys.\  {\bf 61}, 1 (1989).
\bibitem{Polchinski:2006gy} 
  J.~Polchinski,
  ``The Cosmological Constant and the String Landscape,''
  hep-th/0603249.
\bibitem{Bousso:2007gp} 
  R.~Bousso,
  ``TASI Lectures on the Cosmological Constant,''
  Gen.\ Rel.\ Grav.\  {\bf 40}, 607 (2008)
  [arXiv:0708.4231 [hep-th]].
\bibitem{Bousso:2004fc} 
  R.~Bousso and J.~Polchinski,
  ``The string theory landscape,''
  Sci.\ Am.\  {\bf 291}, 60 (2004).
\bibitem{Weinberg:1987dv} 
  S.~Weinberg,
  ``Anthropic Bound on the Cosmological Constant,''
  Phys.\ Rev.\ Lett.\  {\bf 59}, 2607 (1987).
\bibitem{Abbott:1984qf} 
  L.~F.~Abbott,
  ``A Mechanism for Reducing the Value of the Cosmological Constant,''
  Phys.\ Lett.\  {\bf 150B}, 427 (1985).
\bibitem{Brown:1988kg} 
  J.~D.~Brown and C.~Teitelboim,
  ``Neutralization of the Cosmological Constant by Membrane Creation,''
  Nucl.\ Phys.\ B {\bf 297}, 787 (1988).







  

\bibitem{Dvali:1998pa} 
  G.~R.~Dvali and S.~H.~H.~Tye,
  ``Brane inflation,''
  Phys.\ Lett.\ B {\bf 450}, 72 (1999)
  [hep-ph/9812483].


\bibitem{Burgess:2001fx} 
  C.~P.~Burgess, M.~Majumdar, D.~Nolte, F.~Quevedo, G.~Rajesh and R.~J.~Zhang,
  ``The Inflationary brane anti-brane universe,''
  JHEP {\bf 0107}, 047 (2001)
  [hep-th/0105204].


\bibitem{Sarangi:2002yt} 
  S.~Sarangi and S.~H.~H.~Tye,
  ``Cosmic string production towards the end of brane inflation,''
  Phys.\ Lett.\ B {\bf 536}, 185 (2002)
  [hep-th/0204074].


\bibitem{Copeland:2003bj} 
  E.~J.~Copeland, R.~C.~Myers and J.~Polchinski,
  ``Cosmic F and D strings,''
  JHEP {\bf 0406}, 013 (2004)
  [hep-th/0312067].


\bibitem{Polchinski:2015bea} 
  J.~Polchinski,
  ``Brane/antibrane dynamics and KKLT stability,''
  arXiv:1509.05710 [hep-th].


\bibitem{Bekenstein:1972tm} 
  J.~D.~Bekenstein,
  ``Black holes and the second law,''
  Lett.\ Nuovo Cim.\  {\bf 4}, 737 (1972).


\bibitem{Bardeen:1973gs} 
  J.~M.~Bardeen, B.~Carter and S.~W.~Hawking,
  ``The Four laws of black hole mechanics,''
  Commun.\ Math.\ Phys.\  {\bf 31}, 161 (1973).


\bibitem{Hawking:1974sw} 
  S.~W.~Hawking,
  ``Particle Creation by Black Holes,''
  Commun.\ Math.\ Phys.\  {\bf 43}, 199 (1975)
  Erratum: [Commun.\ Math.\ Phys.\  {\bf 46}, 206 (1976)].


\bibitem{Hawking:1976ra} 
  S.~W.~Hawking,
  ``Breakdown of Predictability in Gravitational Collapse,''
  Phys.\ Rev.\ D {\bf 14}, 2460 (1976).


\bibitem{Susskind:1993if} 
  L.~Susskind, L.~Thorlacius and J.~Uglum,
  ``The Stretched horizon and black hole complementarity,''
  Phys.\ Rev.\ D {\bf 48}, 3743 (1993)
  [hep-th/9306069].


\bibitem{Almheiri:2012rt} 
  A.~Almheiri, D.~Marolf, J.~Polchinski and J.~Sully,
  ``Black Holes: Complementarity or Firewalls?,''
  JHEP {\bf 1302}, 062 (2013)
  [arXiv:1207.3123 [hep-th]].


\bibitem{Fischler:1990pk} 
  W.~Fischler, D.~Morgan and J.~Polchinski,
  ``Quantization of False Vacuum Bubbles: A Hamiltonian Treatment of Gravitational Tunneling,''
  Phys.\ Rev.\ D {\bf 42}, 4042 (1990).


\bibitem{Kane:1994vb} 
  C.~L.~Kane, M.~P.~A.~Fisher and J.~Polchinski,
  ``Randomness at the edge: Theory of quantum Hall transport at filling nu = 2/3,''
  Phys.\ Rev.\ Lett.\  {\bf 72}, 4129 (1994)
  [cond-mat/9402108].


\bibitem{Thorne:1994xa} 
  K.~S.~Thorne,
  ``Black holes and time warps: Einstein's outrageous legacy,''
  London, UK: Picador (1994) 619 p


\bibitem{books}
  J.~Polchinski,
  "String Theory Vol. 1,2". Cambridge University Press (1998).
  ``String theory. Vol. 1: An introduction to the bosonic string,''


\bibitem{Polchinski:1992ed} 
  J.~Polchinski,
  ``Effective field theory and the Fermi surface,''
  In {\em Boulder 1992, Proceedings, Recent directions in particle theory} 235-274;
  [hep-th/9210046].



\bibitem{Polchinski:1996nb} 
  J.~Polchinski,
  ``String duality: A Colloquium,''
  Rev.\ Mod.\ Phys.\  {\bf 68}, 1245 (1996)
  [hep-th/9607050].



\bibitem{Polchinski:2010hw} 
  J.~Polchinski,
  ``Introduction to Gauge/Gravity Duality,''
  arXiv:1010.6134 [hep-th].


\bibitem{Polchinski:2014mva} 
  J.~Polchinski,
  ``Dualities of Fields and Strings,''
  Stud.\ Hist.\ Phil.\ Sci.\ B {\bf 59}, 6 (2017)
  [arXiv:1412.5704 [hep-th]].

\bibitem{Polchinski:2015pzt}
  J.~Polchinski,
  ``String theory to the rescue,''
  arXiv:1512.02477 [hep-th].
  
  \bibitem{Polchinski:2016xto}
  J.~Polchinski,
  ``Why trust a theory? Some further remarks (part 1),''
  arXiv:1601.06145 [hep-th].




\end{thebibliography}
\end{document}